# Wall-attached structures of velocity fluctuations in a turbulent boundary layer

Jinyul Hwang and Hyung Jin Sung[†]

*Department of Mechanical Engineering, KAIST*
*291 Daehak-ro, Yuseong-gu, Daejeon 34141, Korea*

[†]Correspondence to Hyung Jin Sung

Tel)   82-42-350-3027

Fax)   82-42-350-5027

E-mail)  hjsung@kaist.ac.kr

**ABSTRACT**

Wall turbulence is a ubiquitous phenomenon in nature and engineering application, yet predicting such turbulence is difficult due to its complexity. High-Reynolds-number turbulence, which includes most practical flows, is particularly complicated because of its wide range of scales. Although the attached-eddy hypothesis postulated by Townsend can be used to predict turbulence intensities and serves as a unified theory for the asymptotic behaviors of turbulence, the presence of attached structures has not been confirmed. Here, we demonstrate the logarithmic region of turbulence intensity by identifying wall-attached structures of velocity fluctuations ($u_i$) through direct numerical simulation of a moderate-Reynolds-number boundary layer ($Re_\tau \approx 1000$). The wall-attached structures are self-similar with respect to their heights ($l_y$), and in particular the population density of the streamwise component ($u$) scales inversely with $l_y$, which is reminiscent of the hierarchy of attached eddies. The turbulent intensities contained within the wall-parallel components ($u$ and $w$) exhibit the logarithmic behavior. The tall attached structures ($l_y^+ > 100$) of $u$ are composed of multiple uniform momentum zones (UMZs) with a long streamwise extent, whereas those of the cross-stream components ($v$ and $w$) are relatively short with a comparable width, suggesting the presence of tall vortical structures associated with multiple UMZs. The magnitudes of the near-wall peak observed in the streamwise turbulent intensity increase with increasing $l_y$, reflecting nested hierarchies of the attached $u$ structures. These findings suggest that the identified structures are prime candidates for Townsend's attached-eddy hypothesis and serve as cornerstones for understanding the multiscale phenomena of high-Reynolds-number boundary layers.

**Key words:** direct numerical simulation, turbulent boundary layers, turbulent flows

## 1. Introduction

Understanding wall-bounded turbulent flows is a long-standing challenge because of their complex and chaotic nature. The presence of a wall not only induces the formation of a thin shear layer close to the wall known as the turbulent boundary layer (TBL), where most of the energy consumption occurs in modern vehicles and pipelines but also separates the TBL into different layers composed of multiscale fluid motions. These multiscale phenomena can be characterized in terms of the friction Reynolds number ($Re_\tau = \delta u_\tau/\nu$), which is the ratio of the viscous length scale $\nu/u_\tau$ ($\nu$ is the kinematic viscosity, and $u_\tau$ is the friction velocity) and the flow thickness $\delta$. Although much progress has been achieved in characterizing the onset of turbulence (Hof *et al.* 2004; Avila *et al.* 2011) and fully turbulent flows at low $Re_\tau$ (Kawahara *et al.* 2012), little progress has been made in the case of high $Re_\tau$ turbulence (Smits *et al.* 2011; Jiménez 2012; Barkley *et al.* 2015), which arises in engineering devices and atmospheric winds ($Re_\tau = O(10^{4-6})$), due to the wide range of scales that govern the transport of mass, momentum and heat.

To elucidate these multiscale phenomena, one approach is to examine the organized motions that retain their spatial coherence for relatively long periods, known as eddies or coherent structures, because these structures are responsible for the dynamical mechanisms and turbulence statistics (Robinson 1991; Adrian 2007). The dominant coherent structures in the buffer layer are low-speed streaks and quasi-streamwise vortices (Kline *et al.* 1967) that are generated via a self-sustaining cycle (Hamilton *et al.* 1995). Above the buffer layer, the coherent structures are larger and more complex due to the presence of various scales. In this region, the mean streamwise velocity ($\overline{U}$) follows a logarithmic profile along the wall-normal distance $y$ (Millikan 1938):

$$\overline{U}^+ = \kappa^{-1}\ln(y^+) + A, \qquad (1.1)$$

where $\overline{U}^+ = \overline{U}/u_\tau$, $y^+ = u_\tau y/\nu$, $\kappa$ is the von Kármán constant, $A$ is the addictive constant, and the overbar indicates an ensemble average. The logarithmic profile in (1.1) represents that the only relevant scales in this region are $y$ and $u_\tau$ (i.e. $\partial \overline{U}/\partial y \sim u_\tau/y$). At high Reynolds numbers, most of the bulk production and velocity drop originate from the logarithmic layer (Smits *et al.* 2011; Jiménez 2012). Townsend (1976) deduced a model for energy-containing eddies in the logarithmic layer whose size scales with $y$; these structures are *attached* to the wall and self-similar. By assuming that the logarithmic layer consists of the superposition of the attached eddies and that the variation of the Reynolds shear stress across the layer is small compared to the viscous stress, the turbulence intensities ($\overline{u_i^2}$) are expressed as

$$\overline{u^2}^+ = B_1 - A_1 \ln(y/\delta), \quad (1.2a)$$

$$\overline{w^2}^+ = B_2 - A_2 \ln(y/\delta), \quad (1.2b)$$

$$\overline{v^2}^+ = B_3, \quad (1.2c)$$

where $u$ (= $u_1$), $v$ (= $u_2$) and $w$ (= $u_3$) indicate the streamwise, spanwise, and wall-normal velocity fluctuations, respectively, and $A_j$ ($j = 1, 2$) and $B_i$ are constant; $A_j$ referred to as Townsend–Perry constant that is expected to be universal (Marusic *et al*. 2013). Here, the wall-parallel components follow the logarithmic variation, while the wall-normal component is constant. The inviscid assumption for self-similar eddies leads to the logarithmic behavior in (1.2*a,b*) and the impermeable condition for the wall-normal component gives rise to (1.2*c*). Perry & Chong (1982) extended this hypothesis by introducing hierarchies of geometrically similar eddies with the probability distribution function (PDF) that is inversely proportional to their height. Based on this approach, they derived the logarithmic variation of $\overline{U}$ (1.1) and $\overline{u_i^2}$ (1.2*a,b*) simultaneously in a sense of the attached-eddy hypothesis. Additionally, they predicted the emergence of a $k_x^{-1}$ ($k_x$ is the streamwise wavenumber) region in the spectrum that is the spectral signature of the attached eddies. In this regard, the attached-eddy hypothesis is a unified theory that links the asymptotic behaviors of the turbulence statistics of high-Reynolds-number flows.

Subsequently, several studies (Perry *et al*. 19986; Perry & Marusic 1995; Marusic & Kunkel 2003) have refined the model of Perry & Chong (1982) to test Townsend's hypothesis, but the Reynolds numbers are not sufficiently high enough to establish the logarithmic region. Nickels *et al*. (2005) showed the presence of the $k_x^{-1}$ region. The coexistence of the logarithmic regions for $\overline{U}$ and $\overline{u^2}$ was observed at $Re_\tau = O(10^{4-5})$ (Hultmark *et al*. 2012; Marusic *et al*. 2013). Vallikivi *et al*. (2015), through the spectral analysis over the same range of $Re_\tau$, observed a region of constant energy in the vicinity of the outer peak, which can contribute to the logarithmic variation for $\overline{u^2}$. For the spanwise component, Jiménez & Hoyas (2008) observed the logarithmic variation at $Re_\tau \approx 2000$. Nevertheless, the central question that has not been resolved is as follow: what is the actual structure in the fully turbulent flow that accords with an attached eddy and forms the logarithmic region? Although Townsend (1976) and Perry & Chong (1982) proposed a particular shape of eddies based on the flow visualization, these structures are modeled to formulate the inverse-power-law PDF and the constant shear stress. Additionally, the presence of the $k_x^{-1}$ region does not necessarily indicate the attached structure, because one coherent motion can carry the energy with a broad range of wavenumbers (Nickels & Marusic 2001) and the wavenumber at a given $y$ does not reflect whether that motion is attached to the wall or is part of a detached one (Jiménez 2013).

To overcome these limitations, clusters of vortices (del Álamo *et al*. 2006) and three-dimensional sweeps/ejections (Lozano-Durán *et al*. 2012) were identified in direct numerical simulation (DNS) of channel flows. These structures can be classified as either wall-attached or wall-detached based on the minimum distance from the wall. The former are self-similar and statistically dominant structures in the logarithmic layer. Hwang (2015) showed the self-similar statistical motions with respect to $y$ in a large-eddy simulation, which restricts the spanwise length scale of motions. Using a proper orthogonal decomposition, Hellström *et al*. (2016) found that the azimuthal length scales of the energetic modes are proportional to the distance from the wall. In addition, Baars *et al*. (2017) reported the self-similar region in the linear coherence spectrum where the coherence magnitude is quantified in a single streamwise/wall-normal aspect ratio. Although these identified coherent motions are reminiscent of Townsend's attached eddy in a sense of self-similarity, it has not been shown how they contribute to the formation of the logarithmic behaivor in $\overline{u^2}$ and $\overline{w^2}$.

The objective of the present study is to identify the self-similar coherent structures satisfying the attached-eddy hypothesis from DNS data of zero-pressure-gradient TBL at $Re_\tau \approx 1000$. To do so, we extract the clusters of the velocity fluctuations ($u_i$). We find that a group of $u_i$ clusters over a wide range of scales is attached to the wall and self-similar. In particular, the population density of the attached *u* clusters is inversely proportional to their height. With these attached clusters, we can reconstruct the turbulent intensities from the superposition of the identified structures. The logarithmic behaviors of the wall-parallel components are verified using the indicator function, which has not been shown in the experiments. We further focus on the *u* clusters in order to explore the hierarchical nature of the structures. The wall-normal distribution of the instantaneous streamwise velocity contained within the objects shows step-like jumps, representing zones of roughly constant uniform momentum. In addition, it is shown that the magnitude of the near-wall peak of the streamwise turbulent intensity carried by the attached clusters of *u* exhibits a logarithmic increase with increasing their heights. The present results not only support the attached-eddy hypothesis but also provide direct evidence regarding the presence of the attached structures in the instantaneous flow field, even at the moderate-Reynolds-number TBL.

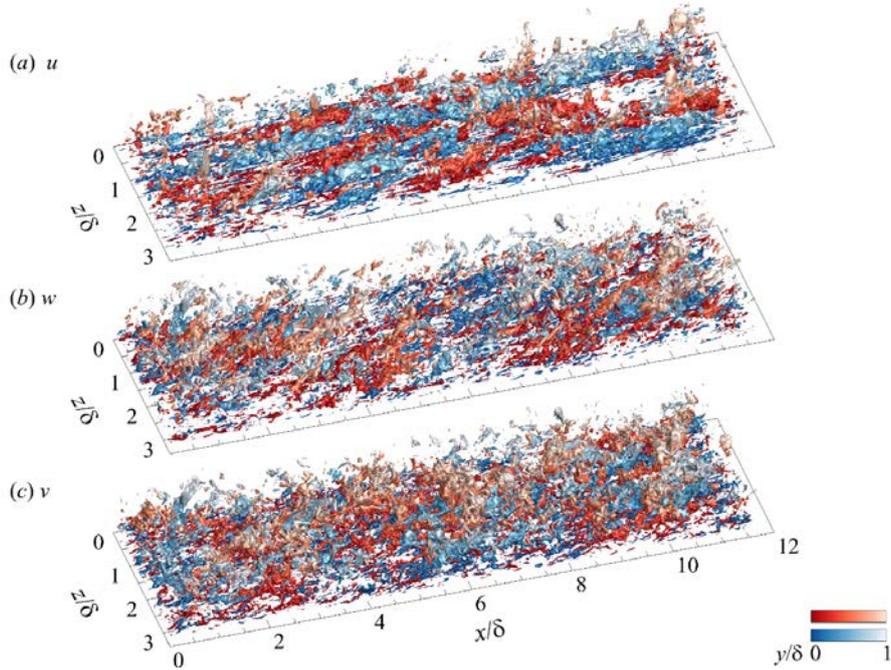

Figure 1. Clusters of velocity fluctuations ($u_i$) at $Re_\tau = 980$. Red and blue isosurfaces indicate the positive and negative fluctuations, $u_i(\boldsymbol{x}) = \pm \alpha u_{rms,i}(y, \delta_t)$, in the instantaneous flow field, respectively: (*a*) *u*; (*b*) *w*; and (*c*) *v*. The brightness of the color indicates the wall distance. Here, the clusters, which cross the edges of the streamwise and spanwise domains, are excluded to represent the size of each cluster completely.

## 2. Computational details

DNS data of the TBL (Hwang & Sung 2017*a*) are used in the present study. The DNS was performed using the fractional step method of Kim *et al.* (2002) to solve the Navier–Stokes equations for incompressible flow. The computational domain is $2{,}300\delta_0 \times 100\delta_0 \times 100\delta_0$, where $\delta_0$ is the inlet boundary layer thickness, in the streamwise ($x$), wall-normal ($y$) and spanwise ($z$) directions, respectively, and the associated components of velocity fluctuations are $u$, $v$ and $w$; details of the DNS and the validation are provided in Hwang & Sung (2017*a*). We used a total of 2430 instantaneous flow fields at $Re_\tau = 980$ with the streamwise length ($L_x$) of $11.7\delta$; the spanwise length ($L_z$) is $3.2\delta$ and $\delta$ is the 99% boundary layer thickness. Across the domain, the Reynolds-number effect is negligible because $Re_\tau$ ranges from 913 to 1039.

In the present study, the fluctuating velocity components are defined by considering the height of the turbulent/non-turbulent interface (TNTI) (Kwon *et al*. 2016). The wall-normal distance of the instantaneous TNTI ($\delta_t$) is defined using the kinetic energy criteria proposed by Chauhan *et al*. (2014). Here, we used all the velocity components to obtain the local turbulent kinetic energy ($\tilde{k}$) in the frame of reference moving with $U_\infty$ over $3 \times 3 \times 3$ grid. We employed the threshold of $\tilde{k} = 0.09$, which is slightly lower than that of Chauhan *et al*. (2014). As a result, the mean of the TNTI height is $\overline{\delta_t} = 0.88\delta$ that is close to the work of Jiménez *et al*. (2010) who defined the TNTI height based on the enstropy. Note that the results in the present work remained qualitatively unchanged regardless of the TNTI threshold because we focus on the structures within the turbulent region only (Hwang & Sung 2017*b*). The streamwise velocity fluctuations based on the Reynolds decomposition are positive when $\delta_t$ is lower than $\delta$. In other words, although these positive-$u$ regions are located in the free-stream region (non-turbulent region), they are regarded as the turbulent regions. We define the velocity fluctuations $u_i = U_i - \overline{U_i}(y, \delta_t)$ to prevent this contamination. Here, $\overline{U_i}(y, \delta_t)$ is the conditional mean velocity, which is a function of the wall-normal distance and the local height of TNTI (Kwon *et al*. 2016).

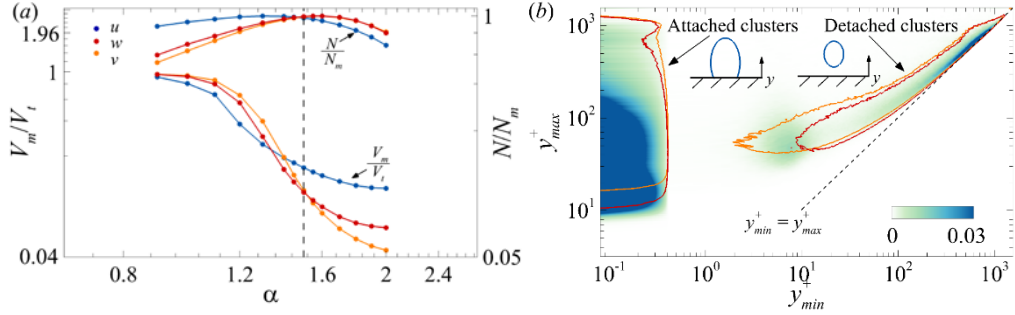

Figure 2. (*a*) Percolation behavior of the identified clusters; the variation of the volume of the largest cluster ($V_m$) normalized by the total volume of the clusters $V_t$, $V_m/V_t$, and ratio of the total number of the clusters ($N$) to the maximum $N$ over $\alpha$ ($N_m$), $N/N_m$. (*b*) The number of the clusters per unit wall-parallel area as a function of $y_{min}$ and $y_{max}$. The color contour indicates the distribution of the $u$ clusters and the red and orange line contours represent the $w$ and $v$ clusters, respectively, with the contour level of 0.003.

The clusters of positive or negative fluctuations are the groups of connected points satisfying

$$u_i(\boldsymbol{x}) > \alpha u_{rms,i}(y,\delta_t) \text{ or } u_i(\boldsymbol{x}) < -\alpha u_{rms,i}(y,\delta_t), \tag{2.1}$$

where $u_{rms,i}$ is the root mean square of the corresponding $u_i$ and $\alpha$ is the threshold. Figure 1 shows the isosurfaces of $u$, $w$ and $v$ in the instantaneous flow field. As seen, the shapes of the structures are complex and some of them are attached to the wall whereas others are distributed far above the wall with a small volume. In particular, the structures of the cross-stream components are pronounced in the edge of the boundary layer; the lighter color denotes the structures whose wall-normal distance is close to $\delta_t$. In addition, we can observe the long meandering structures of $u$ (darker red or blue isosurfaces in figure 1*a*) compared to figure 1(*b*) and (*c*) which will be discussed further in §3. To characterize the irregular shapes of the $u_i$ clusters (figure 1), the connectivity of $u_i$ was defined based on the six orthogonal neighbors of each node in Cartesian coordinates (Moisy & Jiménez 2004; del Álamo *et al.* 2006; Lozano-Durán *et al.* 2012). Using the connectivity rule, the contiguous points are determined at a given node. As a result, we could measure the sizes of clusters and the velocity information over a bounded volume of each object without making an a priori assumption or applying a filter. However, the detection of the structure depends on the threshold value $\alpha$. When $\alpha$ is low, new contiguous points are detected and some of them merge with the previously identified regions. Figure 2(*a*) shows the ratio between the maximum volume of the cluster ($V_m$) and the total volume of the clusters ($V_t$).

Here, $V_m$ and $V_t$ are the sums of the corresponding clusters for negative and positive fluctuations. The black line indicates the total number of the clusters ($N$) normalized by its maximum ($N_m$), which appears at $α_m ≈ 1.4$ for $u$ and $α_m ≈ 1.6$ for $v$ and $w$. At $α > 1.7$, the clusters of intense $u_i$ with very small volume are identified. The volume ratio of $v$ and $w$ clusters converges to 0.05 and 0.07, respectively, representing the very small volume of the intense structures for the cross-stream component; $V_m/V_t ≈ 0.13$ for $u$. As $α$ decreases, $V_m/V_t$ increases significantly and converges to 1. Many new clusters are generated and amalgamated simultaneously, and the amalgamation among the clusters becomes dominant at $α < α_m$, which leads to the decrease of $N/N_m$. This result is consistent with the percolation transition in turbulence structures, such as the vortical structures (Moisy & Jiménez 2004; del Álamo *et al*. 2006) and the ejection/sweep structures (Lozano-Durán *et al*. 2012). The threshold $α = 1.5$ (vertical dashed line) was chosen based on a percolation transition of the clusters; the results remained qualitatively unchanged in the vicinity of the threshold (see Appendix A).

In order to identify the wall-attached structures, the minimum and maximum $y$ ($y_{min}$ and $y_{max}$) of each cluster are measured from the wall. Figure 2(*b*) shows the number of clusters per unit wall-parallel area ($L_x × L_z$) according to $y_{min}$ and $y_{max}$. Here, the color contour indicates the distribution of the $u$ clusters and the inserted line contours represent the distributions of the $w$ (red) and $v$ (orange) clusters. There are two distinct regions, yielding that the clusters are classified into two groups; wall-attached structures with $y_{min}^+ ≈ 0$ and the detached structures with $y_{min}^+ > 0$. The minimum wall-normal distance of the wall-attached group is located at $y_{min}^+ = 0.08$, which is the closest grid point to the wall in the present DNS data. Although previous studies (del Álamo *et al*. 2006, Lozano-Durán *et al*. 2012) classified the clusters of the vortex and Reynolds shear stress into wall-attached and wall-detached groups based on $y_{min}^+ ≈ 20$, we defined the attached structures of the velocity fluctuations with $y_{min}^+ ≈ 0$ because most of the clusters within $y_{min}^+ < 20$ are at $y_{min}^+ ≈ 0$ (88%, 76% and 78% for $u$, $w$ and $v$, respectively). In addition, we can obtain the turbulent intensity carried by the attached structures according to their height ($l_y$) without any interpolation since $l_y = y_{max}$ not $l_y = y_{max} − y_{min}$. In other words, the attached structures identified in the present study physically adhere to the wall. As a result, these structures show the characteristics of the logarithmic layer as well as the buffer layer. This will be discussed further in §4 and 5.

Before addressing the characteristics of the attached structures, it is worth mentioning the term 'attached' used in the attached-eddy hypothesis (Townsend 1976). Townsend conjectured that the main energy-containing motions in the constant shear stress layer (i.e., high Reynolds number) are attached to the wall and thus the eddies are inviscid in this sense. As a result, the slip boundary condition is assumed for the wall-parallel velocity components (i.e., $u$ and $w$) since the hypothesis is valid far above the region where the viscosity is dominant. For the wall-parallel component, Hwang (2016) showed that the energy-containing motions in the form of Townsend's attached eddies penetrate into the region close to the wall (i.e., the footprint as reported in Hutchins & Marusic 2007) and the near-wall penetration of these motions exhibits the inner-scaling behavior. In this respect, the present definition for the attached clusters does not contradict the attached-eddy hypothesis but rather their near-wall parts could contain the viscous effect, which is not considered in the attached-eddy hypothesis (see further §5).

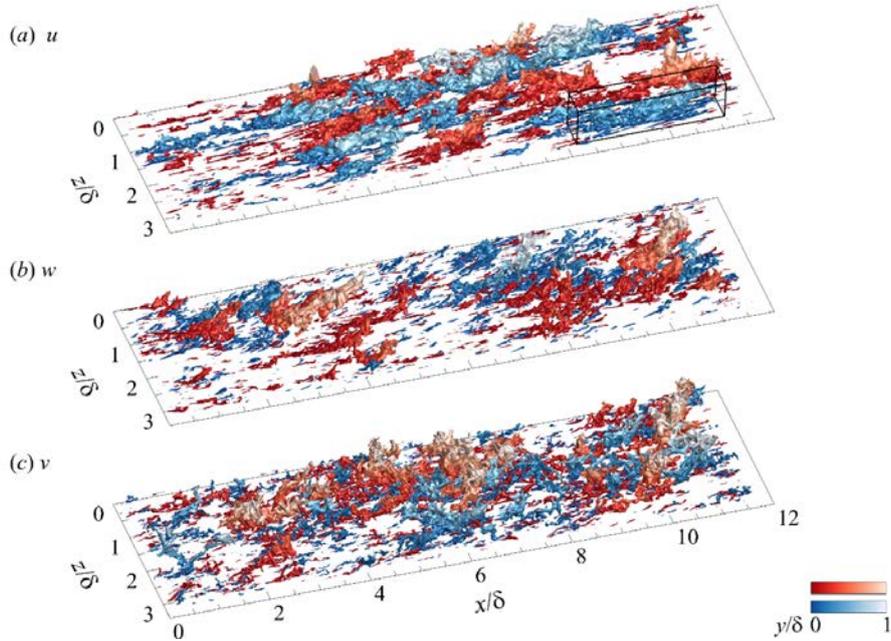

Figure 3. Isosurfaces of wall-attached structures extracted from figure 1: (*a*) *u*; (*b*) *w*; and (*c*) *v*.

|  | Total number | Number fraction | Volume fraction |
|---|---|---|---|
| Attached $u$ | 824396 | 0.20 | 0.67 |
| Detached $u$ | 3215270 | 0.80 | 0.33 |
| Attached $w$ | 1344597 | 0.19 | 0.58 |
| Detached $w$ | 5663332 | 0.81 | 0.42 |
| Attached $v$ | 2353495 | 0.24 | 0.58 |
| Detached $v$ | 7575383 | 0.76 | 0.42 |

Table 1. The number and volume fractions with respect to the total number and total volume of all the identified $u_i$ clusters.

Although the wall-normal velocity is affected by the impermeable condition, which leads to an absence of the logarithmic variation of its intensity profile in a sense of the attached-eddy hypothesis, we can observe the attached structures of $v$ in figure 2(*b*). However, this is not so surprising, given the identification criteria of the present approach (2.1). We used the root mean square of the velocity fluctuations ($u_{rms,i}$), which varies with $y$. Hence, $u_{rms,i}$ reaches zero at the wall, and in particular the value of the wall-normal component is much less than that of the wall-parallel components in the near-wall region. This behavior is also related to the presence of the attached clusters of ejections/sweeps (Lozano-Durán *et al*. 2012) since the Reynolds shear stress is the product of $u$ and $v$.

In contrast to the cross-stream component, there is a weak peak at $y_{min}^+ \approx 7$ and $y_{max}^+ \approx 50$ for the $u$ clusters in figure 2(*b*). These structures can be the fragments of the large attached structures or an object that is developing into the larger one; their number of occurrence is less than 0.011 per unit wall-parallel area. In the present work, we focus on the attached structures whose heights vary from the near-wall region to $\delta$. The number and volume fractions of the identified clusters are summarized in Table 1. Here, we only measure the $u_i$ clusters whose volume is larger than $30^3$ wall units (del Álamo *et al*. 2006, Lozano-Durán *et al*. 2012). Note that more than 90% of the discarded clusters is the detached structure and the rest is the attached one. Although the number fractions are about 20% for the attached objects, they contribute more than half of the total volume of all the clusters. In particular, the attached $u$ structures account for 67% of the volume, representing that these structures tend to exist in large attached clusters compared to the others. Figure 3 illustrates the wall-attached structures of $u_i$ extracted from figure 1. As seen, there are small attached objects close to the wall and very large objects that extend to the edge of the boundary layer with the streamwise elongation.

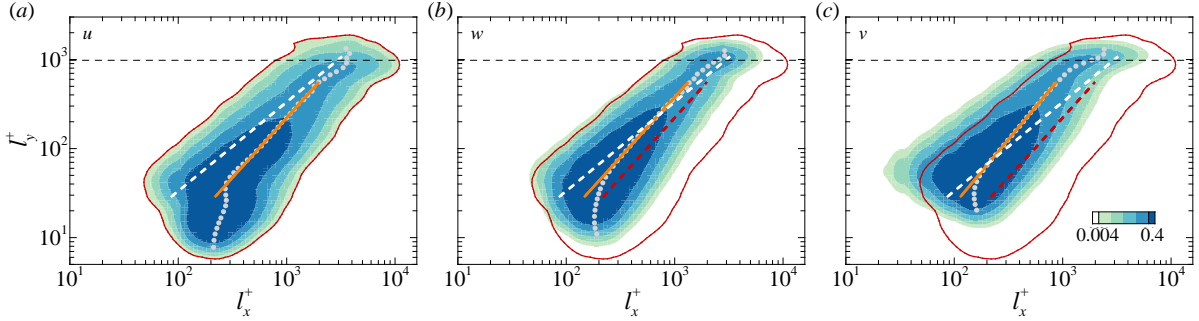

Figure 4. Joint PDFs of the logarithms of the length ($l_x$) of the attached structures and of their height ($l_y$): (a) $u$; (b) $w$; and (c) $v$. The inserted dots indicate the mean $l_x$ with respect to $l_y$. The orange solid line is $l_x^+ = 17.98(l_y^+)^{0.74}$ in (a), $l_x^+ = 12.10(l_y^+)^{0.74}$ in (b), and $l_x^+ = 11.34(l_y^+)^{0.69}$ in (c). In (b) and (c), the red contour line represents the contour level of 0.004 in (a) and the red dashed line is consistent with the orange line in (a). The white dashed line ($l_x^+ = 3l_y^+$) indicates the scaling of the vortex and Reynolds shear stress clusters (del Álamo et al. 2006; Lozano-Durán et al. 2012). The horizontal dashed line indicates $l_y^+ = \delta^+$. The contour levels are logarithmically distributed.

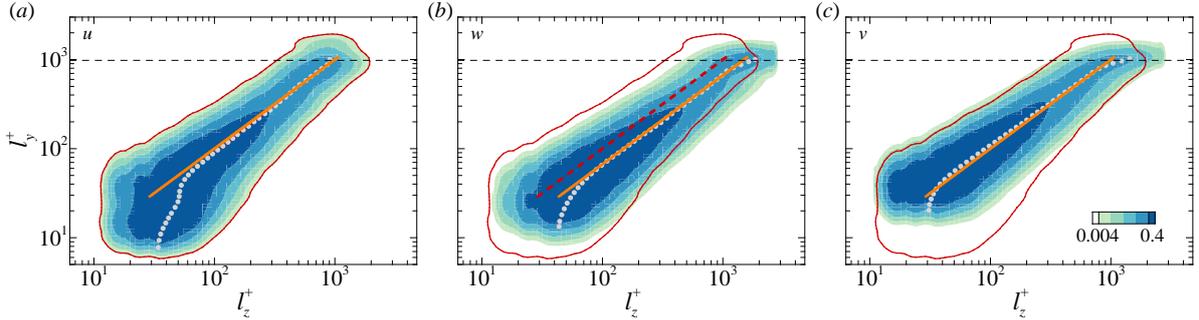

Figure 5. Joint PDFs of the logarithms of the width ($l_z$) of the attached structures and of their height ($l_y$): (a) $u$; (b) $w$; and (c) $v$. The inserted dots indicate the mean $l_z$ with respect to $l_y$. The orange solid line is $l_z^+ = l_y^+$ in (a), $l_z^+ = 1.5l_y^+$ in (b), and $l_z^+ = l_y^+$ in (c). In (b) and (c), the red contour line represents the contour level of 0.004 in (a). The red dashed line in (b) is consistent with the orange line in (a). The horizontal dashed line indicates $l_y^+ = \delta^+$. The contour levels are logarithmically distributed.

## 3. Self-similarity of the attached structures

This section explores the self-similarity of the attached structures with respect to the distance from the wall. The self-similar nature of attached eddies is one of the assumptions of the attached-eddy hypothesis (Townsend 1976). In particular, this assumption leads to the inverse-power-law PDF of hierarchical distribution in Perry & Chong (1982). Hence, the sizes of the identified attached structures and their population density are elucidated to provide evidence for the presence of these structures in the instantaneous flow fields.

### 3.1 *Scaling of the attached structures*

Figures 4 and 5 represent the distributions of the length ($l_x$) and width ($l_z$) of the attached structures with respect to $l_y$, respectively. Here, $l_x$ and $l_z$ are measured based on the bounding box of each structure. There are two distinct growth rates. For the buffer-layer structures ($l_y^+ < 60$), $l_x$ and $l_z$ increase gradually whereas those of the tall structures ($l_y^+ > 100$) grow rapidly until $l_y$ is bounded by $\delta$. For $l_y^+ > 100$, the mean $l_x$ and $l_z$ (circles) scale with $l_y$, representing that the structures are characterized by $l_y$ over a broad range, although there is some dispersion at a given $l_y$. Since the mean $l_x$ and $l_z$ indicate the sizes of representative structures, the dispersion would be associated with attached structures at different stages of stretching (Perry & Chong 1982). Although the mean $l_x$ is not linearly proportional to $l_y$, the mean $l_z$ especially follows a linear law $l_z^+ \approx 1$–$1.5 l_y^+$ (orange lines in figures 4 and 5), indicating that the spanwise length scale of the structures is proportional to the distance from the wall, reminiscent of the attached-eddy hypothesis. This is discussed further below.

As shown in figure 4, the slope of the mean $l_x$ is 0.74 for the wall-parallel components whereas that of the mean $l_x$ is 0.69 for the wall-normal component (orange lines) over $100 < l_y^+ < 550$. In figure 4(*b,c*), the red solid line indicates the distributions of the attached *u* structures and the red dashed lines represent the corresponding orange line shown in figure 4(*a*) for comparison. As $l_y$ increases, the length of the attached *u* and *w* grows with a similar rate beyond the attached *v*: for *v*, the mean $l_x$ is only 860 wall units at $l_y^+ = 550$ ($l_y/\delta \approx 0.55$). In addition, the attached *u* is longer than the transverse components. Given that the width of the attached *u* is comparable with those of the attached *v* and *w* (figure 5), the streamwise organization of the *u* structures contributes to the high volume fraction, 0.67 in Table 1, which is also observed in the instantaneous flow fields (figure 3). To further examine the length distribution, we plot $l_x = 3 l_y$ (white dashed line) which represents the length distribution of the tall attached clusters ($l_y^+ > 100$) of vortex (del Álamo *et al.* 2006) and of ejection/sweep (Lozano-Durán *et al.* 2012) in figure 4. Note that these structures are closely associated with the velocity clusters identified in the present work. Although these works showed the linear relationship between $l_x$ and $l_y$, it should be emphasized that they obtained such a relationship roughly by connecting the ridge of the low contour level. Furthermore, Lozano-Durán *et al*. (2012) pointed out the non-linear relationship between $l_x$ and $l_z$ at a given $l_y$ (i.e. $l_x l_y \propto l_z^2$), which was first observed in the two-dimensional spectra of the streamwise velocity (del Álamo *et al*. 2004), even $l_x$ and $l_z$ vary linearly with respect to $l_y$. However, the attached clusters of $u$ follows $l_x l_y \propto l_z^{1.7}$ that is approximately quadratic, representing that the power law observed in the attached $u$ is suitable for describing the energy-containing motions of the streamwise component. It is also worth to highlight that the distributions of the transverse components are also approximately aligned along $l_x l_y \propto l_z^2$ that was absent in the corresponding energy spectra (del Álamo *et al*. 2004). Recently, Hwang (2015) showed the bimodal behavior of energy-containing motions (i.e., long streaky motions of $u$ and relatively short vortical structures contained all the velocity components) and consequently the scaling of these motions satisfies the quadratic distribution of the energy spectra in del Álamo *et al*. (2004). In this respect, the attached structures of $v$ and $w$ represent the short and tall vortical structures described in Hwang (2015), which are associated with hierarchies of hairpin packets (Adrian *et al*. 2000) or tall attached vortex clusters (del Álamo *et al*. 2006).

The absence of the linear relationship between $l_x$ and $l_y$ could be attributed to i) the preferred azimuthal (or spanwise) inclination of the structures or ii) the meandering nature of the structures or iii) the low Reynolds number of the present data ($Re_\tau \approx 1000$). Baltzer *et al*. (2013) suggested that the relatively shorter $u$ structures are aligned with the preferred azimuthal offset in the turbulent pipe flow, which is consistent with the helix angle of roll cells, and form very long structures. Since the long structures possess roll cells (Hutchins & Marusic 2007*b*; Hwang *et al*. 2016*b*; Krug *et al*. 2017), the absence of the linear relationship between $l_x$ and $l_y$ may arise from the preferred offset of the organized motions. Another possibility is the meandering nature of $u$ regions. In the near-wall region, Jiménez *et al.* (2004) showed a power-law relationship between the length and width of the *u* regions by considering the meandering of the near-wall streaks. Given the meandering of long negative *u* in the logarithmic region (Hutchins & Marusic 2007*a*), the attached *u* structure beyond the buffer layer may follow a power-law relationship between $l_x$ and $l_z$. Finally, the Reynolds number of the present TBL would not be enough to establish the linear relationship ($l_x \propto l_y$) in the self-similar range. Recently, Chandran *et al.* (2017) found the linear relationship in the two-dimensional spectra of the streamwise velocity at $Re_\tau \approx 26000$, whereas the shape of the spectra is aligned along the quadratic form ($l_x l_y \propto l_z^2$) at $Re_\tau = 2400$ that is comparable to del Álamo *et al.* (2004) and Hwang (2015). In the spectra of the high Reynolds number, the lower range of large scales follows the quadratic relationship. On the other hand, the larger scales break away from it and are aligned along $l_x \propto l_z$. The linear growth of the streamwise length scale was also reported in Baars *et al.* (2017) at $Re_\tau = O(10^6)$ using a spectral coherence analysis. In figure 4(*a*), the mean $l_x$ of the structures in $550 < l_y^+ < 750$ follows $l_x = 3.8 l_y$, representing that the larger scales tend towards the linear behavior. If we assume that these tall structures exist within the upper limit of the logarithmic region ($y = 0.15\delta$), then a low Reynolds number limit for the linear relationship is $Re_\tau = 750/0.15 = 5000$. Thus, at $Re_\tau > 5000$, we may observe the attached structures of *u* that follow $l_x \propto l_y$ in the self-similar range consistent with the distribution of their width. Note that we used $l_x$, $l_z$ and $l_y$ instead of the wavelengths and the wall-normal location for convenience because the sizes of the identified objects can approximately represent the latter (del Álamo *et al.* 2006).

Next, the linear relationship between $l_z$ and $l_y$ is further examined (figure 5). In contrast to the length distributions (figure 4), the attached structures of $u_i$ with $l_y^+ > 100$ have a comparable $l_z$ in figure 5. Moreover, the mean $l_z$ of these structures are linearly proportional to $l_y$ up to $l_y^+ = \delta^+$ (horizontal dashed line). Specifically, the mean $l_z$ of the streamwise and wall-normal components is aligned along $l_z = l_y$ and that of the spanwise component behaves in $l_z = 1.5 l_y$, which is slightly wider than the others. The tall attached clusters of the Reynolds shear stress (Lozano-Durán *et al*. 2012) follows $l_z = l_y$ similar to figure 5(*a,c*) since the Reynolds shear stress is the product of *u* and *v*. The widths of the tall vortex clusters (del Álamo *et al*. 2006) are aligned along $l_z = 1.5l_y$ consistent with those of the attached *w*, representing that the spanwise scale of large-scale vortical structures is governed by the attached structures of *w* with a similar $l_y$. In addition, it is worth mentioning that there is a linear ridge, which connects the inner and outer peaks in the spanwise spectra of the streamwise velocity (Hwang 2015). Since the linear ridge represents the self-similarity of the energy-containing motions over a broad range, Hwang (2015) found that the energy distributions in the spanwise spectra of all the velocity, which resolve turbulent motions at a given spanwise length scale, are characterized by the spanwise length scale. Similar behavior is also found in Hellström *et al*. (2016) by examining the spanwise scale of the energetic modes obtained from a proper orthogonal decomposition. It should be emphasized that the self-similar motions found in the previous studies are statistical structures, which partially satisfy the concept of self-similar representative eddies as originally proposed by Townsend who tried to describe correlation statistics of wall turbulence (Marusic *et al*. 2017). However, the present result shows that the existence of self-similar structures in instantaneous flow fields. Further evidence on the nature of the identified attached structures in the context of the attached-eddy hypothesis is explored by examining their population density and the associated turbulent intensity profiles in the remaining part of the present work.

Before getting into this analysis, we need to point out the protrusions of the distributions ($l_y \geq \delta$) in figures 4 and 5 that are also observed in Lozano-Durán *et al*. (2012). Note that we discarded the structures which cross the streamwise domain ($L_x = 11.7\delta$), whereas the objects that cut through the spanwise domain ($L_x = 3.2\delta$) were included. The attached structures of *u* within the protrusions exhibit $l_x = 7$–$11\delta$ and $l_z = 1.5$–$2\delta$ and those of the cross-stream components are $l_x = 4$–$6\delta$ and $l_z = 2$–$3\delta$ that are shorter but wider than the former. These features are consistent with the bimodal behavior of the energy-containing motions at the largest spanwise length scale (Hwang 2015). However, it should be stressed that the largest structures in figures 4 and 5 are not aligned along the orange line (i.e. the scaling failure with respect to $l_y$), although Hwang (2015) described that these motions are self-similar based on their spanwise length scale. Interestingly, Perry *et al.* (1986) conjectured the presence of such large-scale eddies whose heights are an order of $\delta$ and also noted that these eddies need not be self-similar to the smaller-scale eddies. In this respect, the tallest structures ($l_y \approx \delta$) within the scaling failure region are the large-scale eddies, which are not geometrically self-similar with $l_y$. Here, Perry *et al.* (1986) described the tall motions as large scale since large-scale motions were defined originally by their heights such as bulges at the TNTI (Kovasznay *et al.* 1970, Falco 1977). Recently, the streamwise length has been used to characterize large-scale structures and in this sense very long attached structures of $u$ are associated with superstructures in TBL (Hutchins & Marusic 2007*a*); very-large-scale motions (Kim & Adrian 1999) or global modes (del Álamo *et al.* 2004) in internal flows. Hence, a better nomenclature for the attached structures of $u$ with $l_y \approx \delta$ and with relatively shorter $l_y$ (100 < $l_y^+$ < 550) could be superstructures (or very-large-scale motions) and large-scale motions, respectively, but we prefer to use the classical description; i.e. the structures with $l_y \approx \delta$ as the large-scale eddies based on their heights for our purpose.

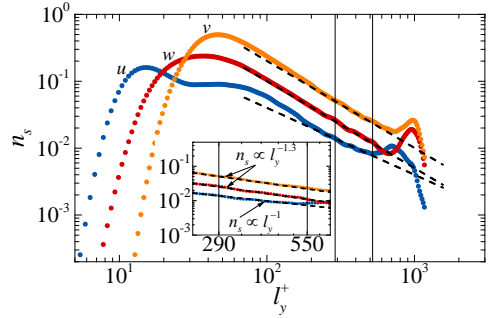

Figure 6. Population density of the attached clusters ($n_s$) with respect to their height $l_y$. The dashed line is $n_s \propto l_y^{-1}$ for $u$ and $n_s \propto l_y^{-1.3}$ for $v$ and $w$. The inset shows a magnified view of the region in 290 < $l_y^+$ < 550.

### 3.2 *Population density of the attached structures*

The population density of the attached structures ($n_s$) versus $l_y$ is examined to determine whether the attached structures are associated the hierarchy length scales (Perry & Chong 1982; Perry *et al.* 1986). Here, $n_{s,i}(l_y)$ is defined

$$n_{s,i}(l_y) = \frac{N_i(l_y)}{mL_x L_z}, \tag{3.1}$$

where $N_i(l_y)$ is the number of the attached structure at a given $l_y$, $m$ is the number of instantaneous flow fields. Note that $n_s$ can be obtained by integrating the joint PDFs in figure 4 or 5 along the abscissa. Hence, $n_s$ is an equivalent measure of the PDF of the hierarchy scales (Perry & Chong 1982). In figure 6, the distributions decay with $l_y$ beyond the buffer layer. In particular, the distribution of the attached $u$ structure (blue) is inversely proportional to $l_y$ while that of the cross-stream components follows $n_s \propto l_y^{-1.3}$ for $290 < l_y^+ < 550$. Given the inverse-power-law PDF of hierarchy scales (Perry & Chong 1982), the structures of $u$ in this region are hierarchies of self-similar eddies in the context of the attached-eddy hypothesis (see further discussion in §5). The population density of $v$ and $w$ are larger than that of $u$ over a broad range consistent with the number of the attached clusters in Table 1. The cross-stream components have the same slope and decrease rapidly compared to $u$, indicating that the attached structures of $u$ exist with larger clusters in instantaneous flow fields. Given that the attached structures of $v$ and $w$ are associated with tall vortical structures as discussed in §3.1, this result also reveals that the attached structures of the cross-stream components and of $u$ exhibit different characteristics. Furthermore, we can observe a peak at $l_y \approx 0.7$–$0.8\delta$ for $u$ and at $l_y \approx 0.9\delta$ for $v$ and $w$, indicating that additional weighting for the large-scale structures. As discussed in §3.1, these large-scale structures are not geometrically self-similar in connection with the protrusions around $l_y \approx \delta$ in figures 4 and 5. This behavior is consistent with the modified PDF of hierarchy scales (Perry *et al.* 1986), which was proposed to enable the more accurate prediction of the mean velocity defect and the low-wavenumber peak in energy spectra.

## 4. Turbulence intensities

The attached structures of $u_i$ identified in the present study show that the self-similar behavior for the objects with their height $l_y$. In addition, their population densities are characterized by $l_y$, and in particular the $u$ structures exhibit the inverse-power-law distribution reminiscent of hierarchies of attached eddies (Perry & Chong 1982). The question then arises: do these attached structures actually form the logarithmic variation of $\overline{u^2}$ as predicted by Townsend (1976)? To answer this question, the turbulent intensities carried by attached structures with different heights $\overline{u_{a,i}^2}(y,l_y)$ are defined as

$$\overline{u_{a,i}^2}(y,l_y) = \left\langle \frac{1}{S_{a,i}(y,l_y)} \int_{S_{a,i}} u_i(x)u_i(x)dxdz \right\rangle, \qquad (4.1)$$

where $S_{a,i}(y, l_y)$ is the wall-parallel area of the structures with $l_y$ at a given $y$ and the angle brackets $\langle \cdot \rangle$ denote an ensemble average. Note that $\overline{u_{a,i}^2}$ indicates the conditionally averaged statistics only for $u_i$ within the structures, not the total turbulent intensities $\overline{u_i^2}$. As a result, the magnitude of $\overline{u_{a,i}^2}$ is larger than that of $\overline{u_i^2}$ because we conditionally average only the intense $u_i$ of the extracted structures (i.e. $|u_i| > \alpha u_{rms,i}$).

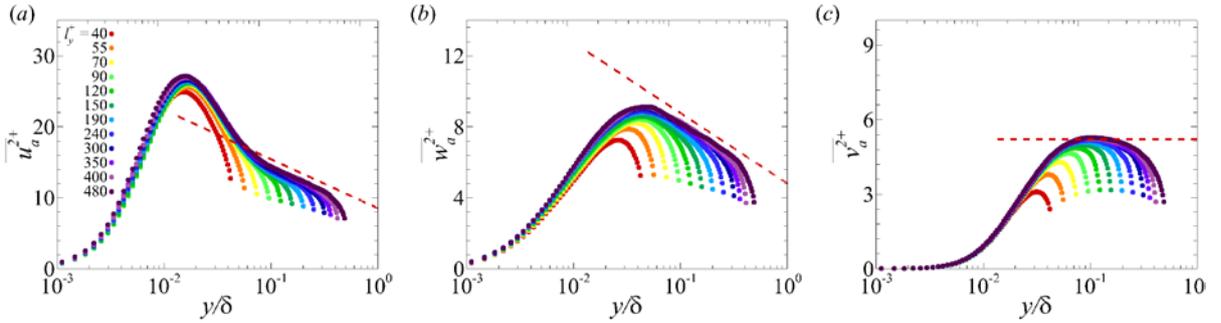

Figure 7. Wall-normal variations of the conditionally averaged turbulent intensity within the attached structures for various $l_y$; (a) $\overline{u_a^2}$; (b) $\overline{w_a^2}$; and (c) $\overline{v_a^2}$. The dashed lines in (a,b) corresponding to the logarithmic variation is a guide for the eye.

Figure 7 shows the wall-normal variation of $\overline{u_{a,i}^2}(y,l_y)$ for various $l_y$. The inserted dashed lines in figure 7(a,b) represent the logarithmic variation. As $l_y$ increases, the profiles of $\overline{u_a^2}$ and $\overline{w_a^2}$ are close to the dashed lines with the emergence of the logarithmic variation. This result is remarkable considering the Reynolds number of the present TBL ($Re_\tau \approx 1000$); the logarithmic behavior of $\overline{u^2}$ and $\overline{w^2}$ was observed at $Re_\tau = O(10^{4-5})$ in experiments (Hultmark et al. 2012; Marusic et al. 2013) and at $Re_\tau = 2000$ in DNS (Jiménez & Hoyas 2008), respectively. On the other hand, the profiles of $\overline{v_a^2}$ (figure 7c) show the plateau and its range extends with increasing $l_y$ consistent with the wall-normal turbulent intensity predicted by Townsend (1976). In addition, the magnitudes of $\overline{v_a^2}$ (figure 7c) are very close to zero at $y/\delta < 0.03$ whereas those of the wall-parallel components (figure 7a,b) are non-negligible at the same range. This behavior is similar to the eddy intensity function of Townsend (1976). However, it should not be confused with the eddy intensity function given that the magnitudes of $\overline{u_a^2}$ and $\overline{w_a^2}$ increase with increasing $l_y$ along $y$ and that there is the near-wall peak (see further discussion in §5.2). As seen in figure 7(a,b), the profiles of $\overline{u_a^2}$ and $\overline{w_a^2}$ do not collapse over a broad range. The profiles in the region where the logarithmic variation appears shift upwards with increasing $l_y$, representing that the addictive constant in (1.2) would depend on the Reynolds numbers; a similar behavior is also found in Jiménez & Hoyas (2008) by assessing the turbulent intensity data in a wide range of $Re_\tau$. In addition, the magnitudes of the near-wall peak increase simultaneously (see figure 10 for details). In contrast to the wall-parallel components, the profiles of $\overline{v_a^2}$ collapse well at $y/\delta < 0.02$. In other words, the structures of $v$ with $y_{min} \approx 0$ do not carry the Reynolds stress and they are topologically attached to the wall due to the criteria (2.1) as discussed in §2. In this sense, these results do not contradict the idea of Townsend (1976).

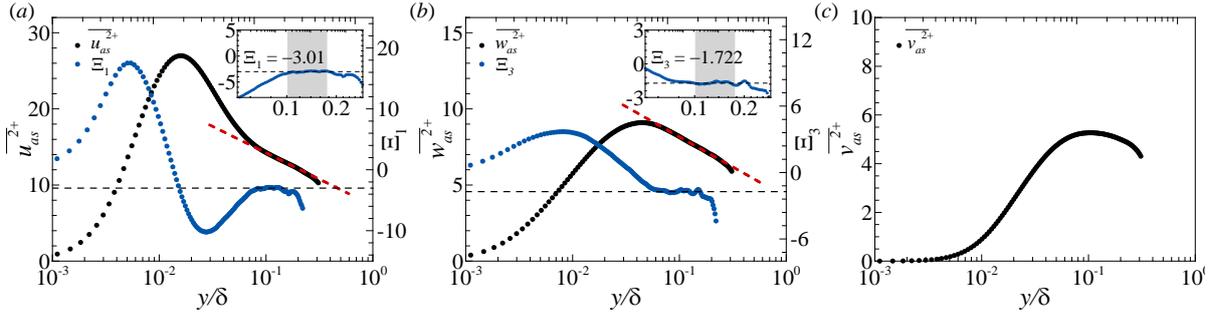

Figure 8. Superposition of the conditionally averaged turbulent intensity carried by the attached structures in $290 < l_y^+ < 550$: (a) $\overline{u_{as}^2}$; (b) $\overline{w_{as}^2}$; and (c) $\overline{v_{as}^2}$. The blue circle indicates $\Xi_1 = y\partial \overline{u_{as}^2}/\partial y$ (a) and $\Xi_3 = y\partial \overline{w_{as}^2}/\partial y \, \overline{u_{as}^2}$ (b), which is the indicator function of the logarithmic law. The inset shows a magnified view of the region where there is a plateau in the range $100 < y^+ < 0.18\delta^+$ (shaded region). The red dashed line correspond to $\overline{u_{as}^2}^+ = 7.3 - 3.01\ln(y/\delta)$ in (a) and $\overline{w_{as}^2} = 4.2 - 1.722\ln(y/\delta)$ in (b).

To further examine the logarithmic behavior in figure 7, we reconstruct the turbulent intensity through the superposition of over $290 < l_y^+ < 550$ where the population density ($n_s$) scales with $l_y$ in figure 6. Especially, $n_s$ of the attached $u$ is inversely proportional to $l_y$. Here, the wall-normal profile of the reconstructed turbulent intensity ($\overline{u_{as,i}^2}$) was computed by weighting the relative probability of the structures to the corresponding $\overline{u_{a,i}^2}$:

$$\overline{u_{as,i}^2} = \frac{\sum_{l_y} n_{s,i}(l_y)\overline{u_{a,i}^2}(y,l_y)}{\sum_{l_y} n_{s,i}(l_y)}. \tag{4.2}$$

The black dot in figure 8 shows the wall-normal variation of $\overline{u_{as,i}^2}$. To confirm the logarithmic variation of $\overline{u_{as}^2}$ and $\overline{w_{as}^2}$, the indicator function

$$\Xi_i(y) = y\partial \overline{u_{as,i}^2}^+ / \partial y, \tag{4.3}$$

which is constant in the logarithmic region, is also plotted in figure 8(*a,b*). For the wall-parallel components, a plateau appears in the same region $100 < y^+ < 0.18\delta^+$ (see shaded region in the inset), verifying the presence of the logarithmic region formed by the attached structures. On the other hand, it is unclear that there is a constant region over $100 < y^+ < 0.18\delta^+$ in the profile of $\overline{v_{as}^2}$ (figure 8*c*). It is worth highlighting that we verify the logarithmic behavior using the indicator function, which was absent in the experiments (Hultmark *et al.* 2012, Marusic *et al.* 2013) that conducted in high-Reynolds number flows due to the experimental uncertainty of the measurements. In addition, although the profile of the spanwise turbulent intensity exhibits the logarithmic variation, this behavior is absent in the streamwise component at $Re_\tau = 2000$ in Jiménez & Hoyas (2008). They noted that the absence of the logarithmic variation is not only due to the viscous effect but also the very long and wide motions (i.e. global modes or very-large-scale motions) since the streamwise turbulent intensity is much closer to the logarithmic variation after removing these motions. Given the fact that the profile of $\overline{u_{as}^2}$ is obtained among the structures with $290 < l_y^+ < 550$ (i.e. $\overline{u_{as}^2}$ does not include the large-scale structures in the protrusions in figures 4 and 5), the present result supports this argument. Overall, the present results not only aid the attached-eddy hypothesis but also provide direct evidence regarding the presence of the attached structures, even in the moderate-Reynolds number TBL.

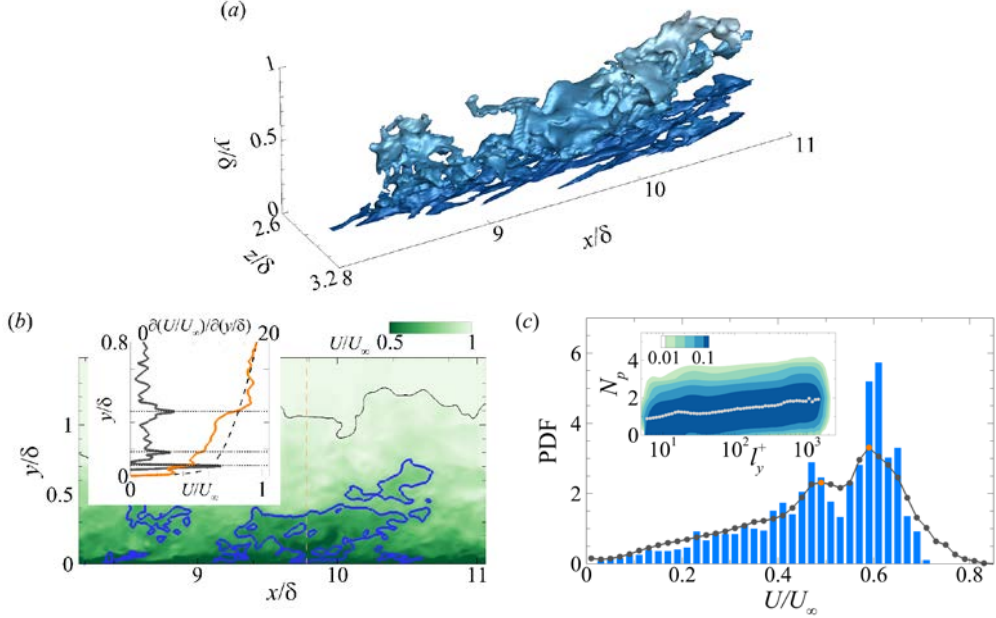

Figure 9. (*a*) A sample attached structure of negative *u* enclosed by the black box in figure 3(*a*). (*b*) Contour of the instantaneous streamwise velocity (*U*) in the *x*–*y* plane at $z/\delta = 2.86$. The blue line is a slice of the object shown in (*a*). The black line indicates the instantaneous TNTI. The inset shows a comparison of the instantaneous (orange line) and mean (dashed line) streamwise velocity profiles at $x/\delta = 9.77$ (indicated by the vertical dashed line in the contour). The grey line represents $\partial U/\partial y$ except the region for $y^+ < 50$ due to the strong shear close to the wall. The horizontal dotted lines indicate step-like jumps of *U* across the structure, which separate the zones of roughly constant *U*. (*b*) Histogram of *U* in the cross-stream plane of the identified object at $x/\delta = 9.77$, which contains the two distinct local maxima (at $U/U_\infty \approx 0.5$ and 0.6) that are associated with possible UMZs. The grey line shows the PDF of *U* within the identified object and the orange circles denote the local maxima. The inset shows the joint PDF of the height ($l_y$) and the number of UMZs ($N_p$). The inserted dots indicate the mean $N_p$ with respect to $l_y$. The contour levels are logarithmically distributed.

## 5. Hierarchies of the attached structures

This section further explores the hierarchical nature of the attached structures, especially for the streamwise velocity fluctuations that exhibit the inverse-power-law distribution (figure 6) as well as the self-similarity (figures 4 and 5) with respect to $l_y$. In addition, long *u* structures play an important role in wall turbulence because long negative-*u* regions are associated with the net Reynolds shear force (Hwang *et al*. 2016*a*), and because the outer

negative-*u* structures extend to the wall and interact with near-wall streaks during the merging of the outer structures (Hwang *et al*. 2016*b*).

### 5.1 *Uniform momentum zones in the attached structures*

As conjectured by Perry & Chong (1982), the attached eddies are in the form of hierarchies. Each hierarchy is composed of the eddies whose height grows from their initial roll-up height ($O(v/u_\tau)$) to the height of the hierarchy (*l*). Here, the height of the hierarchy corresponds to the height of the highest eddy within the hierarchy. In other words, there are several eddies whose heights are less than *l* at a given hierarchy height *l* and consequently the number of eddies or hierarchies increases with increasing *l*. In addition, if we assume that the height of the highest hierarchy is the boundary layer thickness *δ*, the friction Reynolds number ($Re_\tau = \delta u_\tau/v$) indicates the ratio of the highest hierarchy to the initial roll-up height. As a result, the number of eddies or hierarchies can be expected to increase with increasing $Re_\tau$. In this regard, de Silva *et al*. (2016) argued that the log-linear increase in the number of uniform momentum zones (UMZs) with increasing $Re_\tau$ indicates a hierarchical length-scale distribution. This interpretation was further supported by synthesizing the flow field using the attached-eddy model of Maruisc (2001). UMZs that contain roughly uniform streamwise velocity were first observed in the *x–y* plane of instantaneous flow fields by Meinhart & Adrian (1995). In addition, UMZs commonly exist in multiple zones along the wall-normal direction (Adrian *et al*. 2000), implying the hierarchical nature. In order to further examine the hierarchical characteristics of the attached *u* structure, we examine the instantaneous streamwise velocity (*U*) within the object.

Figure 9(*a*) shows a sample attached structure of negative-*u* denoted by the black box in figure 3(*a*) and the streamwise slice at $z/\delta = 2.86$ is illustrated in figure 9(*b*). The sample structure extends from the wall to $y \approx 0.8\delta$ and in particular, the profile of *U* at $x/\delta = 9.77$ (inset in figure 9*b*) show several jumps in velocity across the structure (indicated by dotted horizontal lines), separating zones of roughly uniform *U*; low $\partial U/\partial y$ (orange line) also appears within the UMZs. Given that the UMZs produce local maxima in the histogram of *U* (Adrian *et al*. 2000), we plot the histogram of *U* in the cross-stream plane of the identified structures at $x/\delta = 9.77$ in figure 9(*c*). Although there are several local maxima, the two at $U/U_\infty \approx 0.5$ and 0.6, which are the consequence of UMZs, are preserved when the data are accumulated over the entire structure (grey line). To further examine the number of UMZs in the attached structures, the joint PDF of the height ($l_y$) and the number of local maxima ($N_p$) is shown in the inset of figure 9(*c*). Here, the inserted dots indicate the mean $N_p$ at a given $l_y$. The number of UMZs in each structure increases with increasing $l_y$, representing the hierarchical nature of the identified structures (de Silva *et al*. 2016). In other words, $l_y$ could be consistent with the hierarchy length scale in connection with the results in §3.

In the contrast to the previous work, we analyze the number of UMZs within the attached structures of *u*. At a single Reynolds number (i.e. $Re_\tau = 980$), the number of UMZs logarithmically increase with increasing $l_y$ (figure 9), directly indicating the relationship between the multiple UMZs phenomena (Meinhart & Adrian 1995; Adrian *et al*. 2000) and the hierarchical distribution of the attached structures of *u*. Hence, one attached structure of *u* might be consistent with multiple hairpin packets of various ages and sizes since small packets can be covered by larger packets which convect faster than the smaller ones (Adrian *et al*. 2000); note that the inclination angle of the *u* structures ranges from 8.8° to 16°, which is similar to that of hairpin packets. In addition, given the wake of attached vortex clusters identified in del Álamo *et al*. (2006) is associated with multiple UMZs, the attached structures of *u* might be shrouded in the vortex clusters. However, we need more extensive examination of the three-dimensional vortices surrounding the *u* structures.

One can raise a question about the detection method for the UMZs within the *u* structures; the local maxima observed in the histogram (figure 9*c*) is just due to the conditional sampling of *U* in a selective volume. As first observed in the *x*–*y* plane of instantaneous flow fields by Meinhart & Adrian (1995), the phenomenon of roughly constant uniform momentum is the local event which is random and time-varying. In order to quantify the UMZs, Adrian *et al*. (2000) analyzed the histogram of *U* with the streamwise domain length of 2000 wall units. de Silva *et al*. (2016) also used the similar length to detect the UMZs. In other words, these previous studies sampled *U* selectively in the short streamwise domain at a certain spanwise location. As introduced in Adrian *et al.* (2000), the local maxima in the histogram depend on the time averaging and the streamwise length of the data. This behavior was discussed in Kwon *et al.* (2014) who examined the variation of the local maxima as a function of the streamwise domain length; when the streamwise domain increases up to $6h$ (where $h$ is the channel half-height), several peaks disappear and only one peak survives. To overcome this limitation, the edges of the UMZs should be carefully detected. In the present work, we defined the boundary of the UMZs based on the three-dimensional $u$ structures. As a result, the histogram of $U$ obtained as a function of the height of each $u$ structure shows several maxima even the streamwise length of the structures increases (figure 9*c*). Negative-$u$ regions in the two-dimensional plane represent the UMZs induced by the surrounding vortices (Tomkins & Adrian 2003; Ganapathisubramani *et al.* 2003; Hutchins *et al.* 2005). Retrograde vortices can exist on the shear layer of positive-$u$ structures (Ganapathisubramani *et al.* 2012), which may induce the roughly uniform momentum greater than the mean velocity. In this respect, sampling the streamwise velocity within the $u$ structures is an appropriate method to detect the UMZs in three-dimensional flow fields.

It is worth highlighting the importance of UMZs in a sense of the attached-eddy hypothesis. The work of Meinhart & Adrian (1995) not only showed the general existence of UMZs but also contained that the UMZs are the connection between the discrete system and the continuous system (Perry & Chong 1982). Meinhart & Adrian (1995) used the term 'zones' to emphasize that step-like jumps in $U$ are the instantaneous phenomena, in contrast to 'layers' such as the buffer layer and logarithmic layer that are defined in terms of the mean quantities (section 5.3 in Adrian *et al.* 2000), and concluded that "*The zonal structure reported here offers some new ways of conceptualizing the structure of wall turbulence. The conventional decomposition of the boundary layer into a wall layer, a logarithmic layer and a wake region is based on time-averaged properties of regions of fixed vertical dimension. The zones, in contrast, are instantaneous, time-evolving entities, whose characteristics contribute to the mean properties of each of the convection layers.*". We believe that UMZs are strong evidence for describing the continuous system (Perry &

Chong 1982), which is a result of the randomness in the discrete system. Perry & Chong (1982) assumed that quantum-jump phenomena in the discrete system can be smoothed due to the randomness and jitter of turbulence, leading to the inverse-power-law PDF of the hierarchy length scale in the continuous system. The present findings directly show that the attached *u* structures composed of multiple UMZs follow the hierarchical length scale distribution (figure 9) and contribute to the formation of the logarithmic region (figure 8).

### 5.2 *Near-wall peaks of the streamwise turbulent intensity*

The logarithmic variation of the streamwise turbulent intensity (1.2) implies that the magnitude of $\overline{u^2}^+$ at a fixed $y^+$ is proportional to $\ln(Re_\tau)$ (Jiménez & Hoyas 2008; Jiménez 2012, Marusic *et al*. 2017); the near-wall peak $\overline{u^2_{\max}}^+$, which appears at $y^+ \approx 15$, exhibit $\overline{u^2_{\max}}^+ \sim \ln(Re_\tau)$. This behavior was first observed experimentally in DeGraaf & Eaton (2000) and they suggested a mixed scaling (i.e. $u_\tau$ and $U_\infty$) for $\overline{u^2_{\max}}^+$ in a sense of the attached-eddy hypothesis. Subsequently, several experimental and numerical works reported the logarithmic increase of $\overline{u^2_{\max}}^+$ (Marusic & Kunkel 2003; Hoyas & Jiménez 2006; Sillero *et al*. 2013; Lee & Moser 2015; Ahn *et al*. 2015; Örlü *et al*. 2017). At extremely high Reynolds numbers ($Re_\tau > 30000$), some works (Hultmark *et al*. 2012; Vallikivi *et al*. 2015) found the absence of the growth of inner peak but this might be due to the spatial and temporal resolution issues (Sillero *et al*. 2013; Marusic *et al*. 2017). Given that the behavior of $\overline{u^2_{\max}}^+ \sim \ln(Re_\tau)$ can be explained in the context of Townsend's attached-eddy hypothesis, we further examine the near-wall peak of $\overline{u^2_a}^+$ versus $l_y$ (figure 7) in this section because $l_y$ represents the hierarchy length scale.

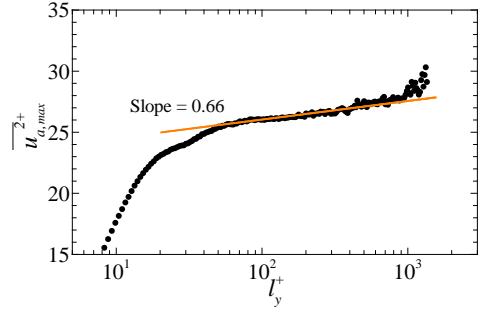

Figure 10. Variation of the near peak magnitude of $\overline{u_a^2}^+$ ($\overline{u_{a,max}^2}^+$) with respect to $l_y^+$. The orange line shows $\overline{u_{a,max}^2}^+ \sim 0.66\ln(l_y^+)$ over $100 < l_y^+ < 550$.

As pointed out in figure 7, the profiles of $\overline{u_a^2}^+$ have a near-wall peak and its magnitude increases with increasing $l_y$. Figure 10(a) shows the variations of the peak $\overline{u_a^2}^+$ according to $l_y^+$. As seen, there is a logarithmic increase of the peak $\overline{u_a^2}^+$ with increasing $l_y^+$ in $100 < l_y^+ < 550$ with a slope of 0.66. This result is in good agreement with the results for the slope of the increase in the peak $\overline{u^2}$ versus $Re_\tau$ obtained in recent DNSs 0.65 (Sillero *et al.* 2013) and in experiments 0.63 (Marusic *et al.* 2017). Sillero *et al.* (2013) and Marusic *et al.* (2017) noted that this value is approximately half of $A_1$ (i.e. the slope of the logarithmic term in 1.2a) since the lower bound of the logarithmic region scales with $Re_\tau^{0.5}$ in contrast to the classical scaling (e.g. $O(100\nu/u_\tau)$ in Perry & Chong (1986). As discussed in §4, the profile of $\overline{u_a^2}^+$ is the result of the conditional averaging of the intense $u$ within the identified structures ($|u| > \alpha u_{rms}$), leading to the difference in $A_1$; the magnitude of $A_1$ for $\overline{u_a^2}^+$ is larger than that for $\overline{u^2}$ ($A_1 = 1.26$ in Marusic *et al.* 2013). However, the slope of the peak $\overline{u_a^2}^+$ is similar to the total turbulent intensity, representing that the attached structures of $u$ could play a dominant role in the increase of the near-wall peak for the total turbulent intensity. However, the Reynolds number effect on the attached structures is required to clarify this argument. The TBL is composed of the attached, detached and weak-$u$ structures ($|u_i| < \alpha u_{rms,i}$). At the present Reynolds number, the attached structures contribute to 20–35% of $\overline{u^2}$ at $y < 0.3\delta$, while the

contributions of the other components are 75–80% leading to the absence of the logarithmic behavior in $\overline{u^2}$ over $100 < y^+ < 0.18\delta^+$. In addition, it is instructive to examine the Reynolds number dependence of the lower bound of the logarithmic region for $\overline{u_a^2}$, which is consistent with the onset of the self-similarity in figures 4 and 5 (i.e. $l_y^+ = 100 \approx 3Re_\tau^{0.5}$ in the present data), but this is beyond the scope of the present work.

Finally, it is worth highlighting that individual attached structures of $u$ do not correspond to a single hierarchy even though $l_y$ is analogous to the height of the hierarchy. As shown in figure 7(a), the magnitude of $\overline{u_a^2}(y, l_y)$ increase with increasing $l_y$ along $y$ and there is the near-wall peak at $y^+ \approx 15$. In other words, the attached structure of $u$ at a given $l_y$ could contain the hierarchies, whose height is less than $l_y$, as well as the process of roll-up which is smaller than that first hierarchy (see figures 18 and 20 in Perry & Chong 1982). In this regard, the profiles of $\overline{u_a^2}$ do not correspond to Townsend eddy intensity functions widely used in various attached-eddy models. Townsend conjectured the distribution of the eddy intensity functions that attached eddies are independent of the viscosity and are self-similar with distance from the wall. On the other hand, the profiles of $\overline{u_a^2}$ are obtained in fully resolved DNS data. As a result, we can observe the logarithmic increase of the peak $\overline{u_a^2}^+$ with $l_y^+$. In other words, the profile of $\overline{u_a^2}^+$ represent the collective contribution of the attached $u$ structures whose heights are less than a given $l_y$. One might think that this physical interpretation contradicts the inverse-power-law PDF, which is inversely proportional to the height of the hierarchy. Since the jitter and randomness of the discrete system or the continuous distribution of hierarchies lead to the inverse-power-law PDF as discussed in Perry & Chong (1982), it is not surprising that the present attached structures follow $1/l_y$; this concept is also related to the percolation behavior of turbulent structures (§2).

## 6. Conclusions

We have demonstrated for the first time that the wall-attached structure of $u$ are energy-containing motions satisfying the attached-eddy hypothesis (Townsend 1976), not only because they are self-similar to $l_y$, but also because there are two strong pieces of evidence: (i) the inverse-power-law PDF, and (ii) the logarithmic variation of the streamwise turbulent intensity ($\overline{u_a^2}$) carried by the identified structures. In particular, we show the presence of the logarithmic region by reconstructing the intensity profile from the superposition of the wall-attached structures in spite of the absence of the logarithmic behavior in the total turbulent intensity at the present Reynolds number ($Re_\tau \approx 1000$). In addition, the wall-attached structures of the cross-stream components (*w* and *v*) exhibit the self-similarity with respect to $l_y$. Although their length is relatively shorter than that of *u*, the widths of all the $u_i$ structures are comparable and linearly proportional to $l_y$ over a broad range, indicating the presence of tall vortical structures. In addition, the spanwise turbulent intensity contained within the attached structures of *w* follows the logarithmic variation over the same range of the streamwise component ($100 < y^+ < 0.18\delta^+$). We further explore the hierarchical natures of the attached *u* structures which exhibit the hierarchical length-scale distribution (PDF ~ $l_y^{-1}$). The wall-normal profile of the instantaneous streamwise velocity within the attached *u* structures shows step-like jumps, reminiscent of UMZs, and in particular the number of UMZs within the objects increases with increasing $l_y$, representing that the structures of *u* are composed of multiple UMZs (or nested hierarchies in a sense of hairpin packet paradigm). Furthermore, the magnitudes of $\overline{u_a^2}$ increase with increasing $l_y$, and especially those of the near-wall peak are proportional to $\ln(l_y^+)$, highlighting that $\overline{u_a^2}$ represents the collective contribution of the attached *u* structures whose heights are less than a given $l_y$. Although we identified the attached structures in a TBL for a single Reynolds number, their hierarchical features will ensure their presence in high-Reynolds-number flows. We anticipate that examining the Reynolds-number effects on attached structures will improve the predictive model (Marusic *et al*. 2010) and exploring their dynamics will facilitate deeper insights into the multiscale energy cascade of wall turbulence.

**Acknowledgements**

This work was supported by the National Research Foundation of Korea (No. 2018001483), and partially supported by the Supercomputing Center (KISTI).

Appendix A. Effect of the structure-identification threshold

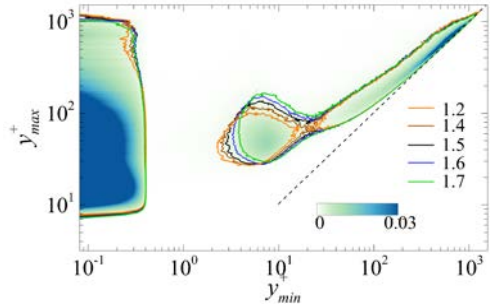

Figure 11. The percolation threshold effect on the number of the $u$ clusters per unit wall-parallel area as a function of $y_{min}$ and $y_{max}$. The color contour is consistent with figure 2(*b*). The line contour is 0.0016 when $\alpha$ = 1.3 (orange), 1.4 (brown), 1.5 (black), 1.6 (blue) and 1.7 (green).

To address the influence of the threshold value $\alpha$, we plot the population density, size distributions, and reconstructed turbulent intensity over a range of threshold from 1.4 to 1.7 where the percolation transition occurs (figure 2*a*). Here, we only include the result of the $u$ clusters to avoid any repetition. Figure 11 shows the threshold effect on the population density of clusters according to $y_{min}$ and $y_{max}$. Regardless of $\alpha$, two distinct regions are observed; the wall-attached and detached groups. As discussed in §2, there is a weak peak at $y_{min}^+ \approx 7$ and $y_{max}^+ \approx 50$. The structures in the vicinity of this peak can be the fragments of the large attached structures or an object that is developing into the larger one. Figure 12(*a,b*) illustrates the distribution of the sizes of the attached structure by varying $\alpha$. The contour lines collapse well, indicating that the self-similarity of the structures with respect to their height ($l_y$) is conserved. Figure 12(*c*) represents the threshold effect on the logarithmic behavior of the reconstructed streamwise turbulent intensity $\overline{u_{as}^2}^+$. Although the profile of $\overline{u_{as}^2}^+$ is shifted upwards with increasing $\alpha$, the logarithmic variation is observed with a similar slope at each α. In the inset, it is evident that the indicator function ($\Xi_1$) has a constant value over the region $100 < y^+ < 0.18\delta^+$ (shaded region).

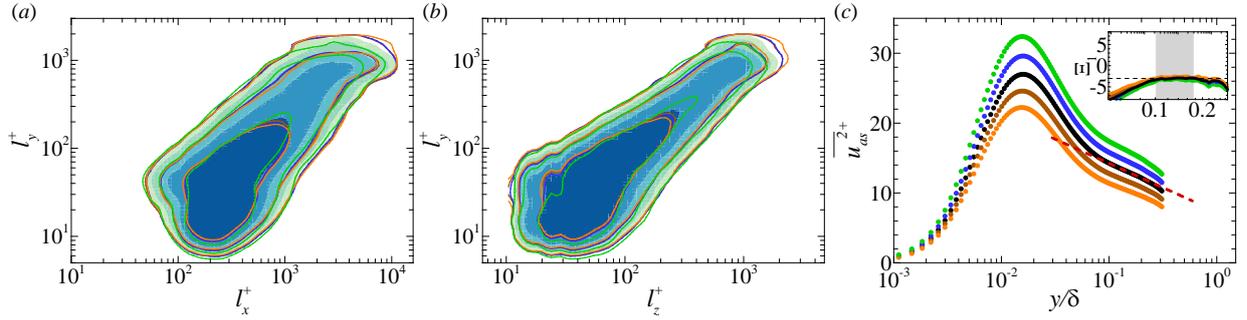

Figure 12. (*a,b*) Joint PDFs of the logarithms of the sizes ($l_x$ and $l_z$) of the attached structures and of the height ($l_y$). The color contour is consistent with figures 4 and 5. The line contour is 0.004, 0.04, and 0.4 with varying α. (*c*) Wall-normal variation of $\overline{u_{as}^2}^+$ with varying α; α = 1.3 (orange), 1.4 (brown), 1.5 (black), 1.6 (blue) and 1.7 (green) The red dashed line indicates the logarithmic variation. The inset shows the corresponding indicator function $\Xi_1$.